\documentclass[aps,preprint,nofootinbib,tightenlines,superscriptaddress]{revtex4}
\usepackage{epsfig}
\usepackage{amsmath}
\begin{document}

\begin{flushright}
BNL-73679-2005-JA \\
RBRC-483
\end{flushright}

\title{Transverse-Momentum-Dependent Gluon Distributions and
Semi-inclusive Processes at Hadron Colliders}

\author{Xiangdong Ji}
\affiliation{Department of Physics, University of Maryland,
College Park, Maryland 20742, USA}
\author{Jian-Ping Ma}
\affiliation{Institute of Theoretical Physics, Academia Sinica,
Beijing, 100080, P. R. China}
\author{Feng Yuan}
\affiliation{RIKEN/BNL Research Center, Building 510A, Brookhaven
National Laboratory, Upton, NY 11973}

\date{\today}
\vspace{0.5in}
\begin{abstract}
We study transverse-momentum-dependent (TMD) gluon distributions
and factorization theorems for the gluon-initiated semi-inclusive
processes at hadron colliders. Gauge-invariant TMD gluon
distributions are defined, and their relations to the integrated
(Feynman) parton distributions are explored when the transverse
momentum is large. Through explicit calculations, soft-collinear
factorization is verified at one-loop order for scalar particle
production. Summation over large double logarithms is made through
solving the Collins-Soper equation. We reproduce the known result
in the limit that the transverse momentum of the scalar is large.
\end{abstract}

\maketitle

\section{Introduction}

In recent years, there has been considerable experimental and
theoretical interest in semi-inclusive hard processes, from which
one hopes to learn important information about the nucleon
structure and non-perturbative dynamics of quantum chromodynamics
(QCD). Rigorous theoretical studies in this direction started with
the classical work on semi-inclusive processes in $e^+e^-$
annihilation by Collins and Soper in \cite{ColSop81}, where a QCD
factorization has been proved, and nonperturbative
transverse-momentum-dependent (TMD) parton distributions and
fragmentation functions were introduced \cite{ColSop81,ColSop81p}.
The approach has been applied to semi-inclusive Drell-Yan
production at hadron colliders in \cite{ColSopSte85} and to
semi-inclusive deep inelastic scattering (SIDIS) in
\cite{MenOlnSop96,NadStuYua00}, in both cases with transverse
momentum much larger than $\Lambda_{\rm QCD}$. Meanwhile,
spin-dependent TMD parton distributions and fragmentation
functions were introduced in various processes, which can generate
polarization asymmetries in semi-inclusive scattering
\cite{RalSop79,{Siv90},Col93,{Anselmino:1994tv},MulTan96}. In the
past few years, gauge properties of the TMD parton distributions
have been investigated
\cite{BroHwaSch02,Col02,{BelJiYua03},BoeMulPij03}, and the result
provided a firm theoretical basis for studying the TMD parton
distributions. More recently, the factorization theorems for the
semi-inclusive deep inelastic scattering (DIS) and Drell-Yan
processes have been re-examined in the context of the
gauge-invariant definitions \cite{JiMaYu04,JiMaYu04p,ColMet04}.

In semi-inclusive DIS and Drell-Yan processes, the quark TMD
distributions produce the dominant contribution, while the
contribution from the gluon distributions is power-suppressed
\cite{JiMaYu04}. However, as a fundamental observable of the
nucleon, the gluon TMD distributions contain unique
non-perturbative information, and contribute to a distinct class
of semi-inclusive
processes\cite{MulRod00,Bur04,BoeVog03,Anselmino04}. For example,
a jet-jet correlation asymmetry in
single-transversely-polarized-hadron-hadron scattering may reveal
the role of the so-called gluon ``Sivers function"
\cite{BoeVog03}. Hence it is essential to have a thorough
theoretical understanding of the gluon TMD distributions and
related factorization theorems for gluon-related semi-inclusive
processes. In general, however, QCD factorization for generic hard
processes (e.g., heavy-flavor and jet production) at hadron
colliders is far more complicated than that for DIS and Drell-Yan
processes \cite{ColSopSte86,ColSopSte89,ColMet04}. As a first
step, we consider here scalar-particle production at hadron
colliders through gluon fusion. We assume that the scalar particle
(e.g. Higgs boson) couples to gluons at the leading order of an
effective theory. All-order QCD corrections are taken into account
in a factorization theorem. This process is similar to Drell-Yan
production in that the observed particles have no final-state
strong interactions, and thus QCD factorization can be studied in
a similar manner. In this paper, we focus on the region of low
transverse momentum where the TMD parton distributions are
relevant.

It is well-known that for a transverse momentum $\Lambda_{\rm
QCD}^2 \ll P_\perp^2\ll Q^2$ where $Q^2$ being some hard scale in
the problem (the scalar particle mass square $M^2$ in the present
case), there exist large logarithms of type
$\alpha_s^m\ln^{2m-1}Q^2/P_\perp^2$ in high-order perturbative
calculations. To have reliable predictions, we must re-sum these
large logarithms
\cite{{DokDyaTro80},{ParPet79},{Ste87},{CatTre89}}. In the present
framework, resummation can be achieved by solving the
Collins-Soper evolution equation for the TMD parton distributions
\cite{ColSop81,ColSopSte85}. We will illustrate how to achieve the
resummation for scalar-particle production at the relevant
transverse momentum.

One interesting example of the scalar particle is the standard
model Higgs boson, whose discovery is among the most important
endeavors for the future high-energy collider experiments. In the
standard model, Higgs boson production at hadron colliders is
dominated by the gluon fusion process: It couples to two gluons
through quark loops, with the dominating contribution from the top
quark loop. This coupling can be described by an effective theory
\cite{Ellis75,{Shifman:1979eb},{Dawson:1990zj},{Djouadi:1991tk}}
This effective theory is valid at the heavy quark limit, and in
practice it works very well at the Higgs mass range of $M_H<2M_t$
\cite{Kramer:1996iq}. In the past few years, there have been
extensive discussions about Higgs production in this effective
theory approach, with the next-to-next-to-leading order
corrections being calculated
\cite{{Harlander:2002wh},{Anastasiou:2002yz}}. Resummation of
large logarithm at low transverse momentum was studied by various
authors in the literature
\cite{{Catani:1988vd},{Hinchliffe:1988ap},{Kauffman:1991jt},{Yuan:1991we},
{deFlorian:2000pr},{Berger:2002ut},{Kulesza:2003wn},{Bozzi:2003jy}}.
In our study, we use the same effective theory for the coupling
between the scalar and gluons. We start with the factorization
analysis of scalar-particle production at small transverse
momentum. For Higgs boson production at large transverse momentum,
we compare our result to the previous calculations with the
resummation effects at one-loop order
\cite{{Catani:1988vd},{Hinchliffe:1988ap},{Kauffman:1991jt},{Yuan:1991we}}.
In our analysis, large logarithms will appear in the TMD parton
distributions, and they are resummed by solving the corresponding
Collins-Soper evolution equation. The resummation coefficient
functions can be calculated directly from the factorization
formula.

This paper is organized as follows. In Sec.II, we discuss TMD
gluon distributions. First, we provide a gauge-invariant
definition of the unpolarized distribution. We then calculate it
to one-loop order in Feynman gauge. We show that the one-loop
result is free of the soft divergence, and obeys the Collins-Soper
evolution equation. In Sec. III, we discuss in detail the
connection between TMD and integrated parton (quark and gluon)
distributions. In Sec. IV, we consider semi-inclusive scalar
particle production at hadron colliders through the gluon fusion
process, where an effective gluon-scalar vertex is introduced. We
calculate the differential cross section and illustrate
factorization at one-loop order. The relevant hard and soft
factors are also calculated. In Sec. V, resummation of large
logarithms is performed by solving the relevant Collins-Soper
equations. We conclude and discuss future prospects in Sec. VI.

\section{Transverse-momentum-dependent gluon distributions}

In this section, we study transverse-momentum-dependent gluon
distribution, which is an important ingredient for scalar-particle
production at hadron colliders. The discussion follows closely our
previous work  on the quark TMD distributions in semi-inclusive
DIS \cite{JiMaYu04}. We first provide a gauge-invariant definition
of the spin-independent TMD gluon distribution. We then calculate
it in perturbation theory to one-loop order, and show that it
obeys the Collins-Soper evolution equation.

We focus the discussion on the unpolarized gluon TMD
distributions. However, the results can easily be generalized to
the polarized TMD gluon distributions
\cite{MulRod00,Bur04,BoeVog03,Anselmino04}, with which one can
study the various polarization asymmetries at hadron colliders. We
leave those studies for future publications.

\subsection{Gauge-Invariant Definition}

According to the previous studies
\cite{ColSop81,ColSop81p,Col02,BelJiYua03,JiMaYu04}, a gauge
invariant, spin-independent TMD gluon distributions can be defined
through the following matrix element
\begin{eqnarray}
xg(x,k_\perp,\mu,x\zeta,\rho)&=&\int\frac{d\xi^-d^2\xi_\perp}{P^+(2\pi)^3}
    e^{-ixP^+\xi^-+i\vec{k}_\perp\cdot \vec\xi_\perp}\nonumber\\
    &&~~\times \frac{\left\langle P|{F_a^+}_\mu(\xi^-,\xi_\perp)
{\cal L}^\dagger_{vab}(\xi^-,\xi_\perp) {\cal L}_{vbc}(0,0_\perp)
F_c^{\mu+}(0)|P \right\rangle}{S(\xi_\perp,\mu,\rho)} \ ,
\end{eqnarray}
where we have included a soft factor following the TMD quark
distributions\cite{JiMaYu04}. In Sec. IIB, we will discuss the
definition of the soft factor and its contribution. In the above
equation $F^{\mu\nu}_a$ is the gluon field strength tensor,
$F_a^{\mu\nu}=\partial ^\mu A_a^\nu -\partial^\nu A_a^\mu-g
f_{abc} A^\mu_b A^\nu_c$. We use the convention for the gauge
coupling $D^\mu=\partial^\mu +igA^\mu$. Light-cone components are
defined as $k^\pm=(k^0\pm k^3)/\sqrt{2}$. Therefore $P^+$ is the
light-cone momentum of the hadron, and $x$ is the momentum
fraction carried by the gluon, while $k_\perp$ is the transverse
momentum. ${\cal L}_v$ is the gauge link from the past,
\begin{eqnarray}
  {\cal L}_v(\xi^-,{\xi_\perp};-\infty) &=&
  P\exp\left(-ig\int_{-\infty}^{0}
    d\lambda v\cdot A(\lambda v + \xi) \right)  \nonumber \\
    && \times
      P\exp\left(-ig\int^{\infty}_{\xi_\perp}
    d\eta_\perp \cdot A_\perp (-\infty, \eta_\perp) \right)\ ,
\end{eqnarray}
where $A^\mu=A^\mu_c t^c$ is the gluon potential in the adjoint
representation, with $t^c_{ab}=-if_{abc}$. Four-vector $v$ is an
off-light-cone vector $v=(v^-,v^+,v_\perp=0)$ where $v^-\gg v^+$.
In principle, the gauge link is best taken along the light-cone
direction. Such a light-cone gauge link, however, introduces the
light-cone singularities \cite{ColSop81,Col02}. The gauge link is
therefore chosen to be slightly off the light cone to regulate
these singularities \cite{Col89}. With a non-light-like $v$, the
TMD distribution depends on a new scalar $\zeta^2=(2v\cdot
P)^2/v^2$. In light-cone coordinates, $\zeta^2=4(P^+)^2 v^-/v^+$
and is proportional to the hadron energy. Energy evolution for the
TMD distributions is governed by the so-called Collins-Soper
equation \cite{ColSop81}. $\mu$ is the standard renormalization
scale. We use the dimensional regularization and modified minimum
subtraction ($\overline {\rm MS}$) to take care of ultra-violate
divergences.

The gauge link above can be derived in the same way as that in the
quark distributions\cite{BelJiYua03}. Its direction is chosen to
start from $-\infty$, indicating its origin from initial-state
interactions present in hadron-hadron scattering. This is similar
to the gauge link in the TMD quark distributions appropriate for
Drell-Yan production \cite{Col02}. [One can, of course, choose a
gauge link pointing to the future to define a similar distribution
relevant, for example, for DIS.] In nonsingular gauges, the
transverse gauge link vanishes because gauge potentials fall off
rapidly at space-time infinity \cite{BelJiYua03}. In a singular
gauge, e.g. the light-cone gauge, gauge potentials do not vanish
at space-time infinity, so we have to include the transverse gauge
link to guarantee gauge invariance \cite{BelJiYua03}. In the
following calculations, for simplicity, we will work in Feynman
gauge in which the transverse gauge link does not contribute.

We are interested in only the leading-twist contribution.
Therefore, we keep only the leading power in $1/\zeta^2$, and take
the $\zeta^2\rightarrow \infty$ limit whenever it is convenient
and there are no singularities.

Feynman rule for the vertex connecting the probing gluon and the
gauge link \cite{ColSop81p} is shown in Fig.~1. The first term
comes from the gauge link, while the second one from the nonlinear
part of the $F^{\mu\nu}$ tensor. In the following calculations, we
present the results for the contributions from these two vertices
as one term and label it as the gauge link contribution.

\begin{figure}[t]
\begin{center}
\includegraphics[height=6cm]{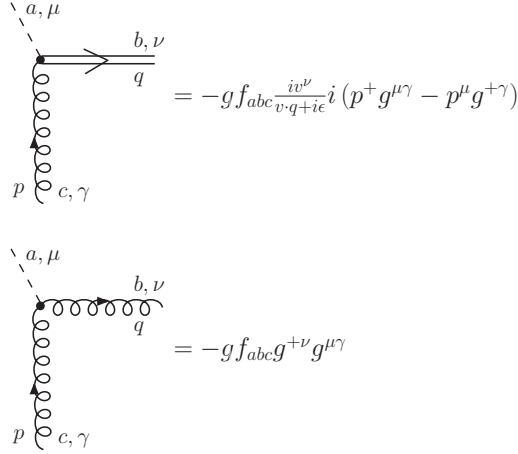}
\end{center}
\vskip -0.7cm \caption{\it The Feynman rule for the TMD vertices.
The first comes from the eikonal line and the second comes from
the non-abelian term in the field strength tensor. In this
calculation, the sum of the two will be lumped together and
labelled by the gauge link.}
\end{figure}

\subsection{One-loop Calculations Without Soft Factor}

In this subsection, we calculate the unpolarized TMD gluon
distribution in a gluon target in perturbative QCD to one-loop
order, ignoring the soft contribution. According to the
definition, the distribution is normalized to
\begin{equation}
g(x,k_\perp)=\delta(x-1)\delta^2(k_\perp) \ ,
\end{equation}
at leading order. At one-loop order, we have both virtual and real
corrections. We regulate the infrared (soft) divergences using
dimensional regularization. Moreover, we introduce the
off-shellness for the initial gluon, $p^2<0$, to regulate possible
collinear divergences.

\begin{figure}[t]
\begin{center}
\includegraphics[height=3.5cm]{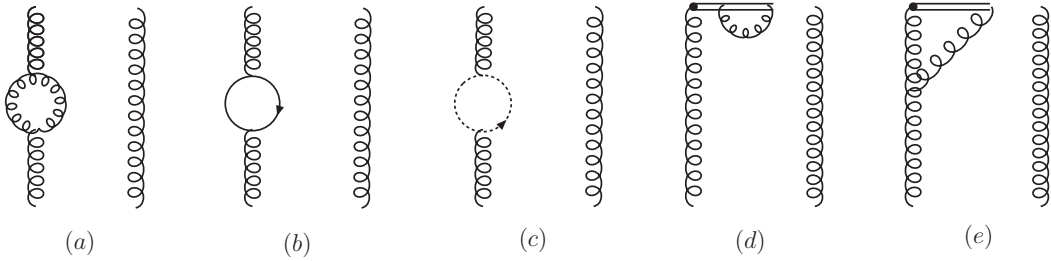}
\end{center}
\vskip -0.7cm \caption{\it Virtual corrections to the TMD gluon
distribution at one-loop order. The mirror diagrams are not
shown.}
\end{figure}

The one-loop virtual diagrams are shown in Fig.~2, which are the
self-energy diagrams for the gluon wave function and from the
gauge link, and the vertex corrections. The contribution from
Figs.~2(a), (b) and (c) has the form
\begin{equation}
g(x,k_\perp)|_{\rm fig.2(a-c)}=\delta (x-1)\delta^2(k_\perp)
(Z_G-1) \ ,
\end{equation}
where
\begin{equation}
Z_G=1+\frac{\alpha_s}{4\pi}\left[-\ln\frac{-p^2}{\mu^2}\left(\frac{5}{3}C_A-
    \frac{2}{3}N_f\right)+\frac{31}{9}C_A-\frac{10}{9}N_f\right ]\
\end{equation}
A factor of 2 has been included to account for the contributions
from the conjugate diagrams. Here $C_A=N_c$, and $N_f$ is the
number of the active quark flavors. Minimal subtraction has been
made to remove the UV divergence. Collinear divergence is
indicated by the dependence on the off-shellness of the gluon.

The contribution from Fig.~2(d) has the similar form, $\delta
(x-1)\delta^2(k_\perp) (Z_W-1)$, with
\begin{equation}
Z_W=1+\frac{\alpha_sC_A}{2\pi}\left[-\frac{2}{\epsilon_{\rm
IR}}+\gamma_E+\ln\frac{1}{4\pi}\right ] \ ,\label{zw}
\end{equation}
where $1/\epsilon_{\rm IR}$ ($\epsilon_{\rm IR}=4-d$) pole comes
from the soft divergence.

The contribution from Fig.~2(e) is $\delta (x-1)\delta^2(k_\perp)
(Z_V-1)$ with
\begin{eqnarray}
Z_V&=&1-\frac{\alpha_sC_A}{2\pi}\left[-\frac{4}{\epsilon_{\rm
IR}^2}-\frac{2}{\epsilon_{\rm IR}}
\left(\ln\frac{\zeta^2}{4\pi\mu^2}-2\ln\frac{-p^2}{4\pi\mu^2}-\gamma_E\right)+
\frac{1}{2}\left(\ln^2\frac{\zeta^2}{4\pi\mu^2}-2\ln^2\frac{-p^2}{4\pi\mu^2}\right)\right. \nonumber\\
    &&~~\left.
+\gamma_E\left(\ln\frac{\zeta^2}{4\pi\mu^2}-2\ln\frac{-p^2}{4\pi\mu^2}\right)
-\ln\frac{\zeta^2}{-p^2}-\frac{1}{2}\ln\frac{\zeta^2}{\mu^2}+
\frac{3}{2}-\frac{\gamma_E^2}{2}+\frac{7\pi^2}{12}\right]
    \ .
\end{eqnarray}
Here we have taken the limit $\zeta^2=(2v\cdot P)^2/v^2$ is much
larger than any soft scale (e.g. $-p^2$). The result shows
explicitly the double logarithmic dependence of the TMD
distribution on $\zeta^2$. These double logarithms can be resummed
using the Collins-Soper evolution equation, which will be
discussed in Sec.~IIE. The total contribution from the virtual
diagrams is
\begin{equation}
g(x,k_\perp,\mu,x\zeta)|_{\rm
virtual}=\delta(x-1)\delta^2(k_\perp)\left(Z_G+Z_W+Z_V-3\right) \
.
\end{equation}
Again the virtual correction contains soft divergences.

\begin{figure}[t]
\begin{center}
\includegraphics[height=3.0cm]{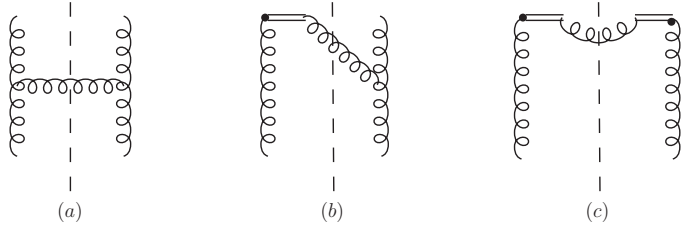}
\end{center}
\vskip -0.7cm \caption{\it Same as Fig.~2 but for the real
contributions. }
\end{figure}

Now we turn to the real corrections shown in Fig.~3. The
contribution from Fig.~3(a) is
\begin{equation}
g(x,k_\perp)|_{\rm
fig.3(a)}=\frac{\alpha_sC_A}{\pi^2}\frac{1}{k_\perp^2-x(1-x)p^2}\left[\frac{1-x}{x}
    +x(1-x)+\frac{x}{2}\right] \ .
\end{equation}
Fig.~3(b) contributes
\begin{equation}
g(x,k_\perp,x\zeta)|_{\rm
fig.3(b)}=\frac{\alpha_sC_A}{\pi^2}\left\{\frac{1}{k_\perp^2-x(1-x)p^2}\left[\frac{x}{(1-x)_+}
-\frac{x}{2}\right]+\frac{1}{2}\frac{\mu^\epsilon}{k_\perp^2}
\ln\frac{\zeta^2}{k_\perp^2}\delta(x-1)\right\}\
 ,
\end{equation}
where a factor of 2 has been included to account for the mirror
diagrams. The plus function follows the definition in
\cite{AltPar77}. The limit of $\zeta^2\rightarrow \infty$ has been
taken whenever there is no divergence. The contribution from
Fig.~3(c) can be calculated similarly,
\begin{equation}
g(x,k_\perp)|_{\rm
fig.3(c)}=-\frac{\alpha_sC_A}{2\pi^2}\frac{\mu^\epsilon}{k_\perp^2}\delta(x-1)
\
\end{equation}
The above contributions contain soft divergences as indicated by
the $1/k_\perp^2$ term as $k_\perp^2\rightarrow 0$. We have to
keep the pre-factors of these terms up to ${\cal O}(\epsilon)$,
which can lead to finite contributions after Fourier transforming
into the impact parameter $b$-space. The soft divergences from the
real and virtual diagrams will eventually cancel, as we shall see.

\subsection{Soft Factor}

As in the case of TMD quark distributions, there are soft
contributions from Figs. 2 and 3, signaled by soft divergences.
These contributions can be extracted using the well-known
Grammer-Yennie approximation \cite{GraYen73}. Here we follow the
same procedure as for the quark distributions \cite{JiMaYu04}, and
define the soft factor
\begin{equation}
S(b,\mu,\rho)=\frac{1}{(N_c^2-1)}\langle 0|{\cal L}_{\bar
vcb'}^\dagger(b_\perp;-\infty) {\cal
L}_{vb'a}^\dagger(-\infty;b_\perp){\cal L}_{vab}(0;-\infty){\cal
L}_{\bar vbc}(-\infty;0)|\rangle \ , \label{softdef}
\end{equation}
where $\bar v$ is another off-light-cone vector, $\bar v=(\bar
v^-,\bar v^+,0)$ with $\bar v^+\gg\bar v^-$, and $\rho$ is defined
as $\rho^2=v^-\bar v^+/v^+\bar v^-$. The same soft factor will
appear in the factorization of semi-inclusive scalar particle
production we shall see in Sec.~IV.

At one-loop order, we have diagrams shown in Fig. 4 which are the
same as those considered in \cite{JiMaYu04}, except for the color
factor. Here we use dimensional regularization for the infrared
divergence instead of a gluon mass. The final result is the same
because the soft factor is free of infrared divergences.

\begin{figure}[t]
\begin{center}
\includegraphics[height=3.0cm]{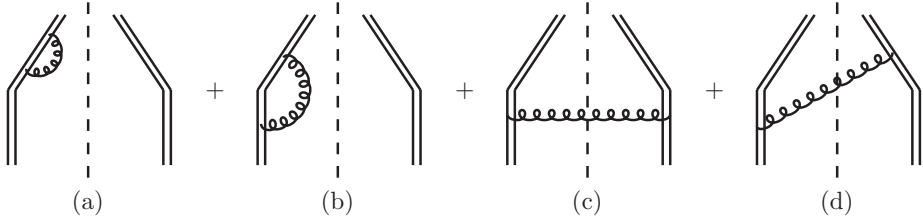}
\end{center}
\vskip -0.7cm \caption{\it Soft contributions to the TMD gluon
distribution at one-loop order. }
\end{figure}

The contribution from Fig.~4(a) is
\begin{equation}
\triangle _{\rm soft} g(x,k_\perp)|_{\rm
fig.4(a)}=-2\delta(x-1)\delta^2(k_\perp) (Z_W-1) \ ,
\end{equation}
where $Z_W$ is in Eq.~(\ref{zw}). The contribution from Fig.~4(b)
is
\begin{equation}
\triangle _{\rm soft} g(x,k_\perp)|_{\rm
fig.4(b)}=\delta(x-1)\delta^2(k_\perp) \frac{\alpha_s
C_A}{2\pi}\ln\rho^2\left(-\frac{2}{\epsilon_{\rm
IR}}+\gamma_E+\ln\frac{1}{4\pi}\right) \ ,
\end{equation}
where the UV divergence has been subtracted using $\overline{\rm
MS}$ scheme. The contribution from Fig.~4(c) is
\begin{equation}
\triangle _{\rm soft} g(x,k_\perp)|_{\rm
fig.4(c)}=\delta(x-1)\frac{\alpha_sC_A}{\pi^2}\frac{\mu^\epsilon}{k_\perp^2}
\ .
\end{equation}
Fig.~4(d) gives,
\begin{equation}
\triangle _{\rm soft} g(x,k_\perp)|_{\rm
fig.4(d)}=-\delta(x-1)\frac{\alpha_sC_A}{2\pi^2}\ln\rho^2\frac{\mu^\epsilon}{k_\perp^2}
\ .
\end{equation}
The individual diagrams all have infrared divergences. However,
they all cancel eventually. For example, the IR divergence from
Fig.~4(a) cancels that from (c), while that in (b) cancels that in
(d). The cancellation can best be seen from the impact parameter
$b$-space expression. After Fourier transformation, we get
\begin{equation}
\triangle _{\rm soft} g(x,b,\mu,x\zeta,\rho)=\delta(x-1)
\frac{\alpha_sC_A}{2\pi}\ln\left(\frac{\mu^2b^2}{4}e^{2\gamma_E}\right)\left[\ln\rho^2-2\right]
\
\end{equation}
We thus confirm that it is the same soft factor as in the quark
distributions \cite{JiMaYu04}, except for the color factor.

We note that there are different definitions of the soft factor in
the literature \cite{JiMaYu04,ColMet04,ColHau00}. The difference
can be viewed as the scheme dependence. Final physical results do
not depend on a particular definition.

\subsection{TMD Gluon Distribution at One-Loop}

Summarizing the results from the last two subsections, we get the
gluon TMD distribution at one-loop order,
\begin{eqnarray}
g(x,k_\perp,\mu,x\zeta,\rho)&=&\delta(x-1)\delta^2(k_\perp)\left[Z_G+Z_V-2+(Z_W-1)(\ln\rho^2-1)\right]\nonumber\\
&&+\delta(x-1)\frac{\alpha_sC_A}{2\pi^2}\frac{\mu^\epsilon}{k_\perp^2}
    \left(\ln\frac{\zeta^2}{k_\perp^2}-\ln\rho^2+1\right)
\nonumber\\
&&
+\frac{\alpha_sC_A}{\pi^2}\frac{1}{k_\perp^2-x(1-x)p^2}\left[\frac{x}{(1-x)_+}+\frac{1-x}{x}+x(1-x)\right]
    \ .
\end{eqnarray}
The individual terms have soft divergences which cancel out in the
sum. In the impact parameter $b$-space, we define
\begin{equation}
g(x,b,\mu,x\zeta)=\int d^2 k_\perp e^{i\vec{k}_\perp\cdot
\vec{b}_\perp} g(x,k_\perp,\mu,x\zeta) \
\end{equation}
After Fourier transformation,
\begin{eqnarray}
g(x,b,\mu,x\zeta)&=&\frac{\alpha_sC_A}{\pi}\left(\ln\frac{4}{-p^2b^2}-2\gamma_E\right)
\left[\frac{x}{(1-x)_+}+\frac{1-x}{x}+x(1-x)\right]\nonumber\\
&&~~+\frac{\alpha_sC_A}{\pi}\left[\left(\frac{x}{1-x}\ln\frac{1}{x(1-x)}\right)_++
\ln\frac{1}{x(1-x)}\left(\frac{1-x}{x}+x(1-x)\right)\right]\nonumber\\
&&~~+\delta(x-1)\left\{\left(Z_G-1\right)+\frac{\alpha_sC_A}{2\pi}\left[
    -\frac{1}{2}\ln^2\left(\frac{\zeta^2b^2}{4}e^{2\gamma_E}\right)
    - \ln
    \frac{-p^2b^2}{4}e^{2\gamma_E}\right.\right.
    \nonumber\\
    &&~~\left.\left.+\ln\frac{\zeta^2b^2}{4}e^{2\gamma_E}
    +\ln\frac{\mu^2b^2}{4}e^{2\gamma_E}\left(
    \ln\rho^2-1\right)+\frac{1}{2}\ln\frac{\zeta^2}{\mu^2}-\frac{\pi^2}{2}-\frac{11}{2}\right]\right\}
    \ .
\end{eqnarray}
The $1/\epsilon_{\rm IR}$ poles have now disappeared. However,
there are still collinear divergences as indicated by the
dependence on $\ln(-p^2)$. More interestingly, we have double
logarithmic dependence on $\zeta^2$, the energy of the parenting
gluon. We study this dependence in the next subsection.

\subsection{Collins-Soper Evolution in Hadron Energy}

From the above result, we see that the TMD gluon distribution
depends on the energy of the parenting hadrons, through the
variable $\zeta^2=(2v\dot P)^2/v^2\approx 2(P^+)^2 v^-/v^+$. The
energy evolution of the TMD parton distribution is controlled by
the Collins-Soper evolution equation \cite{ColSop81}. In
impact-parameter space, the Collins-Soper equation reads as
\begin{equation}
\zeta\frac{\partial}{\partial\zeta}g(x,b,x\zeta,\mu,\rho)=(K_g+G_g)g(x,b,x\zeta,\mu,\rho)
\ ,\
\end{equation}
where $K$ and $G$ are soft and hard evolution kernels,
respectively.

It is easy to check that the one-loop TMD gluon distribution
indeed satisfies the Collins-Soper equation. The sum of $K+G$ can
be extracted,
\begin{equation}
K_g+G_g=-\frac{\alpha_sC_A}{\pi}\ln \frac{x^2\zeta^2
b^2}{4}e^{2\gamma_E-\frac{3}{2}} \ .
\end{equation}
The soft part $K$ can be calculated from a definition similar to
the quark case \cite{Col89, JiMaYu043},
\begin{equation}
K_g=-\frac{\alpha_sC_A}{\pi}\ln\frac{\mu^2 b^2}{4}e^{2\gamma_E} \
,
\end{equation}
and thus the hard part $G_g$ can be solved from the sum. The $K$
and $G$ obey the renormalization group equation,
\begin{equation}
\mu\frac{\partial K_g}{\partial \mu}=-\mu\frac{\partial
G_g}{\partial \mu}=-\gamma_{Kg}=-2\frac{\alpha_sC_A}{\pi} \
\end{equation}
These evolution equations will be used to resum the large
logarithms in the cross section in Sec. V.

\section{TMD Parton Distributions at Large Transverse Momentum}

As emphasized in \cite{JiMaYu04}, it is nontrivial to generate the
integrated (Feynman) parton distributions by integrating out the
transverse momentum in the TMD distributions. The main difficulty
is that new ultraviolet divergences emerge from the transverse
momentum integration. However, there is another way to connect the
two distributions \cite{ColSop81p}: When the transverse momentum
becomes large or the impact parameter becomes small, part of the
TMD distributions are calculable from perturbative QCD, because at
least one hard gluon is needed to generate the large momentum. The
hard-gluon radiation leads to power-like behavior, e.g.
$1/k_\perp^2$, for the unpolarized TMD distributions. Here Feynman
parton distributions enter as the non-perturbative input in the
QCD factorization analysis.

We consider the Fourier-transformed version of TMD distributions
at small $b$. The factorization theorem reads
\cite{ColSop81,ColSop81p},
\begin{eqnarray}
f_i(x,b,\mu,x\zeta,\rho)=\sum_j \int \frac{dy}{y}\tilde
C_{i/j}\left(\frac{x}{y},\mu,\zeta,b,\bar\mu\right) f_j(y,\bar
\mu) \ ,
\end{eqnarray}
where $i,~j$ denote the flavor of the partons---quarks and gluons.
The integrated distributions $f_j$ on the right-hand side depend
on the longitudinal momentum fraction $x$ and the scale $\bar
\mu$. $\tilde C_{i/j}$ are the coefficient functions and are
calculable from perturbative QCD. In the following, we will verify
the above formula and extract the coefficient functions at
one-loop order.

We choose an on-shell quark target (with mass $m$) and an
off-shell gluon target (with off-shellnes $p^2$) to regulate the
collinear divergences. We need to calculate both the TMD and
integrated parton distributions up to one-loop order. The
collinear divergences in both distributions must be matched, and
the subtracted coefficient functions are free of infrared
divergences.

The coefficient functions $\tilde C_{i/j}$ have perturbation
expansions in terms of the strong coupling constant $\alpha_s$. At
leading order ($\alpha_s^0$), it is easy to see,
\begin{eqnarray}
\tilde C_{q/q}^{(0)}&=&\tilde C_{g/g}^{(0)}=\delta (x-1) \ ,
\nonumber\\
\tilde C_{g/q}^{(0)}&=&\tilde C_{q/g}^{(0)}=0 \ .
\end{eqnarray}
At one-loop, all receive non-trivial contributions. $\tilde
C_{q/q}$ has been calculated before in \cite{JiMaYu04}, where a
gluon mass was used to regulate the soft divergences. We have
checked this calculation by using dimensional regularization and
obtained the same result. For completeness, we quote the result
here,
\begin{eqnarray}
  && \tilde C_{q/q}^{(1)}\left(x, b^2, \mu^2, \bar\mu^2,x^2\zeta^2,
\rho\right)
  \nonumber \\
  && = \frac{\alpha_s}{2\pi}C_F\left\{(1-x)+
  \left(\frac{1+x^2}{1-x}\right)_+ \ln
  \frac{4}{b^2\bar\mu^2}e^{-2\gamma_E}   \right.\nonumber \\
&& \left.  +
  \delta(x-1)\left[\left(\frac{1}{2}-\ln\rho^2\right)\ln
\frac{4}{b^2\mu^2}e^{-2\gamma_E}
-\frac{1}{2}\ln^2\left(\frac{\zeta^2b^2}{4}e^{2\gamma_E-1}\right)-\frac{3+\pi^2}{2}
 \right]\right\} \ . \label{ctil}
\end{eqnarray}
To calculate $\tilde C_{g/g}^{(1)}$, we need the integrated gluon
distribution at one loop,
\begin{eqnarray}
g^{(1)}(x,\bar\mu)&=&\frac{\alpha_sC_A}{\pi}\left\{\ln\frac{\bar\mu^2}{-p^2}{\cal
P}_{gg}(x)+\left(\frac{x}{1-x}\ln\frac{1}{x(1-x)}\right)_+-\delta(x-1)
\left(1-\frac{31}{36}+\frac{5N_f}{54}\right)
    \right.\nonumber\\
&&~~~~~~~\left.+\left(\frac{1-x}{x}+
x(1-x)\right)\ln\frac{1}{x(1-x)}\right\}\ ,
\end{eqnarray}
where ${\cal P}_{gg}(x)$ is the gluon splitting function,
\begin{equation}
{\cal P}_{gg}(x)=
    \frac{x}{(1-x)_+}+\frac{1-x}{x}+x(1-x)
    +\delta(x-1)\beta_0\ ,
\end{equation}
with $\beta_0=(11-2/3N_f)/12$. From the above and one-loop TMD
gluon distribution, the coefficient function for the gluon-gluon
term is
\begin{eqnarray}
&&\tilde C_{g/g}^{(1)}\left(x, b^2, \mu^2, \bar\mu^2,x^2\zeta^2,
\rho\right)
\nonumber \\
  && = \frac{\alpha_sC_A}{\pi}\left\{ {\cal P}_{gg}(x)\ln
  \frac{4}{b^2\bar\mu^2}e^{-2\gamma_E}+\delta(x-1)\left[
  \left(\beta_0+\ln\rho-\frac{1}{2}\right)\ln
\frac{b^2\mu^2}{4}e^{2\gamma_E}   \right.\right.\nonumber \\
&& \left.\left. +\frac{3}{4}\ln\frac{\zeta^2}{\mu^2}
     -\frac{1}{4}\ln^2\left(\frac{\zeta^2b^2}{4}e^{2\gamma_E}\right)
     -\frac{\pi^2+7}{4}
 \right]\right\} \ , \label{cggt}
\end{eqnarray}
where the collinear divergences of the form $\ln(-p^2)$ have been
cancelled.

\begin{figure}[t]
\begin{center}
\includegraphics[height=3.0cm]{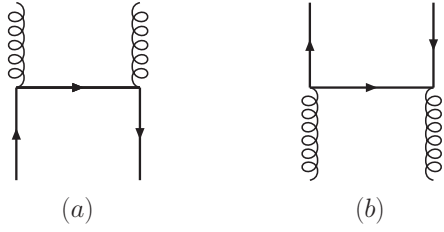}
\end{center}
\vskip -0.7cm \caption{\it Quark splitting to gluon contribution
to $\tilde C_{g/q}$ (a) and gluon splitting to quark $\tilde
C_{q/g}$ (b).}
\end{figure}

The Feynman diagrams for the quark-gluon and gluon-quark splitting
coefficient functions $\tilde C_{g/q}$ and $\tilde C_{q/g}$ are
shown in Fig.~5(a) and (b). The TMD gluon distribution in a quark
target is
\begin{eqnarray}
g(x,k_\perp)|_{\rm
fig.5(a)}=\frac{\alpha_sC_F}{2\pi^2}\left[\frac{1+(1-x)^2}{x}\frac{1}{k_\perp^2+m^2}
-\frac{2x(1-x)m^2}{(k_\perp^2+x^2m^2)^2}\right] \ .
\end{eqnarray}
After Fourier-transforming into the impact parameter $b$-space, we
have,
\begin{eqnarray}
g(x,b)|_{\rm
fig.5(a)}=\frac{\alpha_sC_F}{2\pi}\left[\frac{1+(1-x)^2}{x}\ln
\left(\frac{4}{m^2b^2x^2}e^{-2\gamma_E}\right)
-\frac{2(1-x)}{x}\right] \ .
\end{eqnarray}
The corresponding integrated gluon distribution is,
\begin{eqnarray}
g(x,\bar \mu)|_{\rm
fig.5(a)}=\frac{\alpha_sC_F}{2\pi}\left[\frac{1+(1-x)^2}{x}\ln
\frac{\bar\mu^2}{m^2x^2} -\frac{2(1-x)}{x}-x\right] \ .
\end{eqnarray}
Combining the above, we find the coefficient function,
\begin{eqnarray}
\tilde C_{g/q}^{(1)}\left(x,
b^2,\bar\mu^2\right)=\frac{\alpha_sC_F}{2\pi}\left[\frac{1+(1-x)^2}{x}\ln
\left(\frac{4}{b^2\bar\mu^2}e^{-2\gamma_E}\right) +x\right] \ .
\label{cqg}
\end{eqnarray}
Similarly, we find the coefficient function for finding a quark in
a gluon target,
\begin{eqnarray}
\tilde C_{q/g}^{(1)}\left(x,
b^2,\bar\mu^2\right)=\frac{\alpha_s}{4\pi}\left[\left(x^2+(1-x)^2\right)\ln
\left(\frac{4}{b^2\bar\mu^2}e^{-2\gamma_E}\right) +2x(1-x)\right]
\ .
\end{eqnarray}
These calculations can certainly be carried out to higher orders
in $\alpha_s$.

\section{Scalar-particle production through gluon fusion}

It is difficult to study the gluon TMD distributions in
semi-inclusive DIS because gluons do not have direct couplings
with photons, and their contributions are power suppressed
\cite{JiMaYu04}. However, we can access these distributions at
hadron colliders through, for example, the heavy-quark pair or
di-jet production. In these processes, the total transverse
momentum of the pair or the di-jet provides information on the TMD
parton distributions in the initial hadrons, and the gluon
distributions could dominate at some kinematics.

An important question we have to address is the factorization of
the QCD radiative corrections. Only when the factorization is
established can we safely extract the TMD parton distributions
from data. In this section, we will consider QCD factorization for
the gluon initiated processes, with scalar-particle production as
an example. The scalar particle here has a direct coupling to the
gluons through an effective vertex. The transverse-momentum
distribution of the produced scalars reflect partly the
transverse-momentum dependence of the gluon densities in the
nucleon. Our study can be extended to other gluon initiated
semi-inclusive processes at hadron colliders.

The effective lagrangian for the gluon-scalar coupling is,
\begin{equation}
{\cal L}_{eff}=-\frac{1}{4}g_\phi\Phi F^{a}_{\mu\nu}F^{a\mu\nu} \
,
\end{equation}
where $\Phi$ is the scalar field and $g_\phi$ is the effective
coupling. For the standard model Higgs boson, the effective
coupling can be derived from the full theory
\cite{Ellis75,{Shifman:1979eb}}. The above effective lagrangian
has also been used \cite{Mueller:1989st} to study the gluon
saturation in nuclei.

\begin{figure}[t]
\begin{center}
\includegraphics[height=4.0cm]{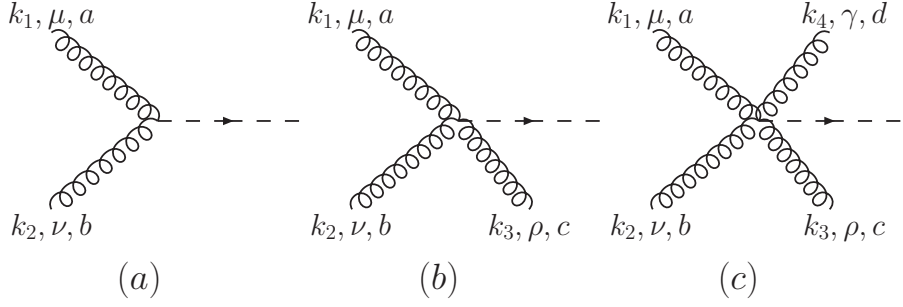}
\end{center}
\vskip -0.7cm \caption{\it Vertices for the scalar particle
coupling to the gluons. All labelled momenta are in.}
\end{figure}

From the above lagrangian, we can write down the basic vertices
for the scalar particle coupling to the gluon potentials, shown in
Fig. 6. The Feynman rule for the coupling with two gluons in
Fig.~6(a) is
\begin{equation}
ig_\phi\delta_{ab}\left(k_1\cdot
k_2g_{\mu\nu}-k_{1\nu}k_{2\mu}\right) \ .
\end{equation}
And the coupling with three gluons in Fig.~6(b) reads
\begin{equation}
-g_\phi g_sf_{abc}\left[(k_1-k_2)_\rho g_{\mu\nu}+(k_2-k_3)_\mu
g_{\rho\nu}+(k_3-k_1)_\nu g_{\rho\mu}\right] \ .
\end{equation}
Finally, the coupling with four gluons shown in Fig.~6(c) is
\begin{eqnarray}
-ig_\phi g_s^2 &\left\{
\right.&f_{abe}f_{cde}\left(g_{\mu\rho}g_{\nu\gamma}-g_{\mu\gamma}g_{\nu\rho}\right)\nonumber\\
&+&f_{ace}f_{bde}\left(g_{\mu\nu}g_{\rho\gamma}-g_{\mu\gamma}g_{\nu\rho}\right)\nonumber\\
&+&\left.f_{ade}f_{bce}\left(g_{\mu\nu}g_{\rho\gamma}-g_{\mu\rho}g_{\nu\gamma}\right)\right\}
\ .
\end{eqnarray}
Production of a single scalar particle has a cross section linear
in $g_\phi^2$. We are interested in higher-order QCD corrections
in $\alpha_s$.

Scalar-particle production at low transverse momentum depends on
the TMD gluon distributions of the incident hadrons. A new feature
for semi-inclusive processes is the dependence on the soft factor
\cite{ColSop81,JiMaYu04}. In a factorized form, the cross section
can be written as a product of the TMD parton distributions (or
fragmentation functions), the soft factor $S$, and the hard factor
$H$ \cite{ColSop81,JiMaYu04},
\begin{eqnarray}
\frac{d^3\sigma(M^2,P_\perp,y)}{d^2P_\perp dy}&=&\sigma_0\int
d^2k_{1\perp}d^2k_{2\perp}d^2\ell_\perp
x_1g(x_1,k_{1\perp},x_1\zeta_1,\mu,\rho)
x_2g(x_2,k_{2\perp},x_2\zeta_2,\mu,\rho)\nonumber\\
&&\times S(\ell_\perp,\mu,\rho)H(M^2,\mu,\rho)
\delta^2(\vec{k}_{1\perp}+\vec{k}_{2\perp}+\vec{\ell}_\perp-\vec{P}_\perp)
\ , \label{fac1}
\end{eqnarray}
where $\sigma_0$ is the leading-order scalar-particle production
from two gluons,
\begin{equation}
\sigma_0=\frac{\pi g_\phi^2}{64}\frac{1}{1-\epsilon/2} \ .
\end{equation}
Here the factor $(1-\epsilon/2)$ in the denominator comes from the
polarization average of the initial gluons. The hard scale $M$
here denotes the mass of the scalar particle, and $y$ and
$P_\perp$ are its rapidity and transverse momentum, respectively.
At low-transverse momentum, the longitudinal-momentum fractions
$x_1$ and $x_2$ for the two incident gluons can be approximately
related to the scalar particle's rapidity $y$ through
$x_1=\sqrt{M^2/S}e^y$ and $x_2=\sqrt{M^2/S}e^{-y}$, where $S$ is
the total center-of-mass energy squared $S=(P_1+P_2)^2$. $\zeta_1$
and $\zeta_2$ are defined as $\zeta_1^2=4(v\cdot P_1)/v^2$ and
$\zeta_2^2=4(\bar v\cdot P_2)/\bar v^2$, and $v$ and $\bar v$ are
defined in the last section. $\rho$ as defined before, is a
parameter to separate gluon contributions to the soft and hard
factors.

The above factorization result is accurate at the leading power in
$P_\perp^2/M^2$ at low transverse momentum. In the following, we
will examine this factorization formula up to one-loop order using
the perturbative TMD gluon distributions from the previous
section. Through explicit computation, we obtain the one-loop
result for the hard factor $H$. For arguments toward a
factorization to all orders, we follow the discussion in
\cite{ColSop81,JiMaYu04}.

In principle, the above factorization formula works only at the
low transverse momentum region, and it breaks down at
$P_{\perp}\sim M$, where one can no longer neglect the power
corrections of $P_\perp^2/M^2$. However, it is useful to
extrapolate the above factorization to all $P_{\perp}$, and
convert it to the impact-parameter space. After Fourier
transformation, the differential cross section can be written as,
\begin{eqnarray}
\frac{d^3\sigma(M^2,P_\perp,y)}{d^2P_\perp
dy}&=&\sigma_0\int\frac{d^2\vec{b}}{(2\pi)^2}e^{-iP_\perp\cdot
b_\perp}W(x_1,x_2,b,M^2) \ ,
\end{eqnarray}
where we define
\begin{eqnarray}
W(x_1,x_2,b,Q^2)= x_1g(x_1,b,x_1\zeta_1,\mu,\rho)
x_2g(x_2,b,x_2\zeta_2,\mu,\rho) S(b,\mu,\rho)H(M^2,\mu,\rho) \
.\label{fac2}
\end{eqnarray}
The convolutions in the transverse-momentum space now reduce to
products in the impact parameter $b$-space.

\subsection{Factorization At One Loop}

In this subsection, we calculate the scalar-particle production to
one-loop order to verify the factorization, and to extract the
hard factor $H$. The gluon TMD distributions and soft factor have
been calculated in Sec.II, where the gluon target has been put
off-shell (by nonzero $-p^2$). In this section, it is more
convenient to demonstrate the factorization by using the on-shell
gluon target ($p^2=0$). It is straightforward to extend the
results in Sec.II to the on-shell case. In the new scheme, the
perturbative TMD gluon distribution is,
\begin{eqnarray}
g(x,b,x\zeta,\mu,\rho)&=&\delta(x-1)\nonumber\\
    &&+\frac{\alpha_sC_A}{\pi}\left\{{\cal P}_{gg}(x)
    \left(-\frac{2}{\epsilon_{\rm IR}}-\gamma_E+\ln\frac{4}{4\pi\mu^2
    b^2}\right)\right.\nonumber\\
&&~~~~\left.+\delta(x-1)\left[\left(\ln\rho+\beta_0-\frac{1}{2}\right)\ln\frac{b^2\mu^2}{4}e^{2\gamma_E}-\frac{1}{4}
    \ln^2\left(\frac{\zeta^2
    b^2}{4}e^{2\gamma_E}\right)\right.\right.\nonumber\\
&&~~~~\left.\left.+\frac{3}{4}\ln\frac{\zeta^2}{\mu^2}-\frac{\pi^2+7}{4}\right]\right\}\
    . \label{tmdgd}
\end{eqnarray}
Here, $1/\epsilon_{\rm IR}$ indicates collinear divergence. The
soft factor $S(b,\mu,\rho)$ was defined in Eq.~(\ref{softdef}). At
one loop, the result in impact parameter space is
\begin{equation}
S(b,\mu,\rho)=\frac{\alpha_sC_A}{2\pi}\ln\left(\frac{b^2\mu^2}{4}e^{2\gamma_E}\right)
\left(2-\ln\rho^2\right) \ . \label{softh}
\end{equation}
These results will be used to verify the above factorization
formula.

\begin{figure}[t]
\begin{center}
\includegraphics[height=3.0cm]{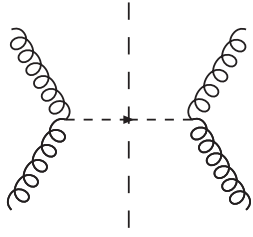}
\end{center}
\vskip -0.7cm \caption{\it Scalar particle production at leading
order.}
\end{figure}

At leading order, the Born diagram for the production cross
section is shown in Fig.7. The calculation is straightforward, and
we get
\begin{equation}
\frac{d\sigma}{d^2P_\perp dy}=\frac{\pi g_\phi^2}{64}\delta
(x_1-1)\delta (x_2-1)\delta^2(P_\perp) \ .
\end{equation}
This leads to
\begin{equation}
W^{(0)}(x_i,b)=\delta(x_1-1)\delta(x_2-1) \ ,
\end{equation}
which can be clearly reproduced by the leading-order gluon
distributions in gluon targets

\begin{figure}[t]
\begin{center}
\includegraphics[height=3.5cm]{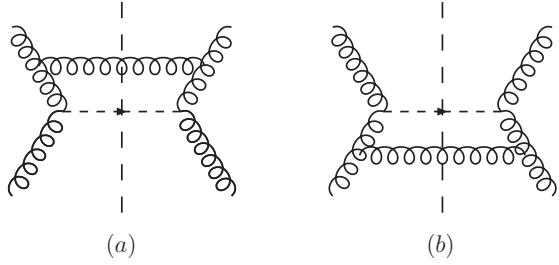}
\end{center}
\vskip -0.7cm \caption{\it Real diagrams contribution to the
scalar particle production through gluon fusion: part I.}
\end{figure}

\begin{figure}[t]
\begin{center}
\includegraphics[height=3.5cm]{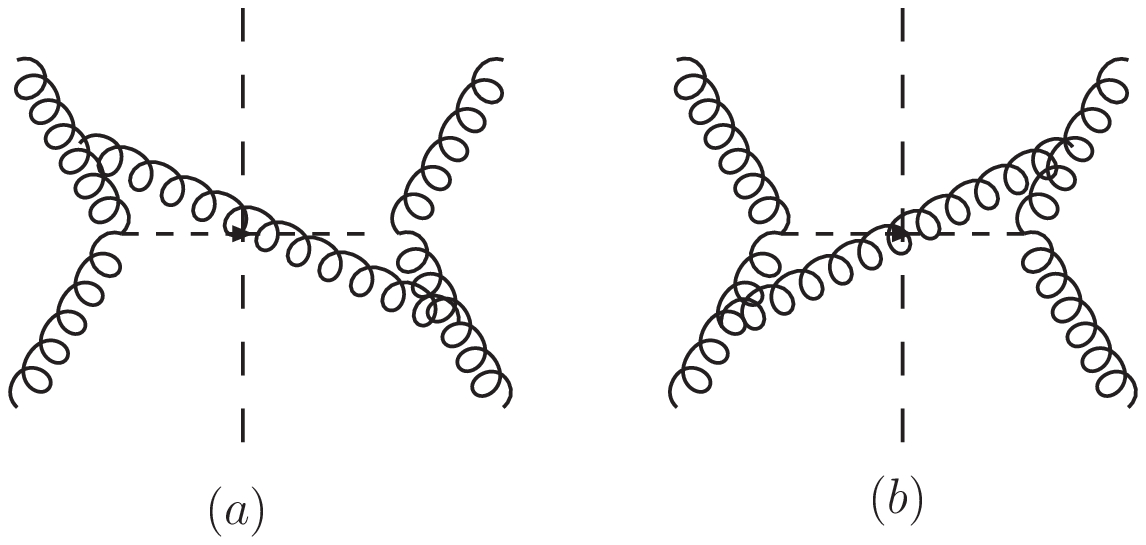}
\end{center}
\vskip -0.7cm \caption{\it Real diagrams: part II.}
\end{figure}

\begin{figure}[t]
\begin{center}
\includegraphics[height=3.5cm]{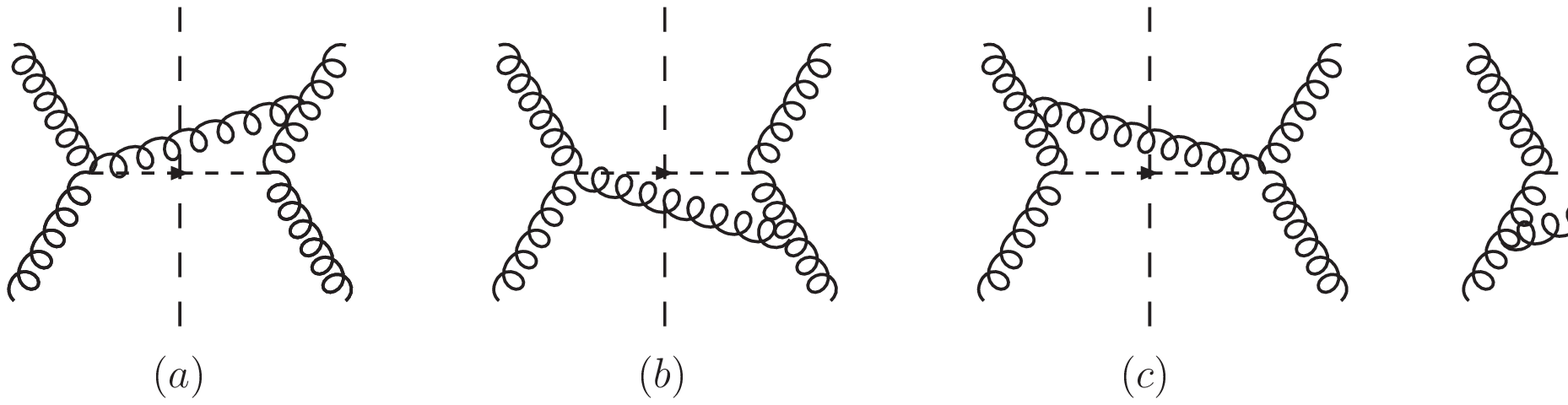}
\end{center}
\vskip -0.7cm \caption{\it Real diagrams: part III.}
\end{figure}

\begin{figure}[t]
\vskip 0.2cm
\begin{center}
\includegraphics[height=6.0cm]{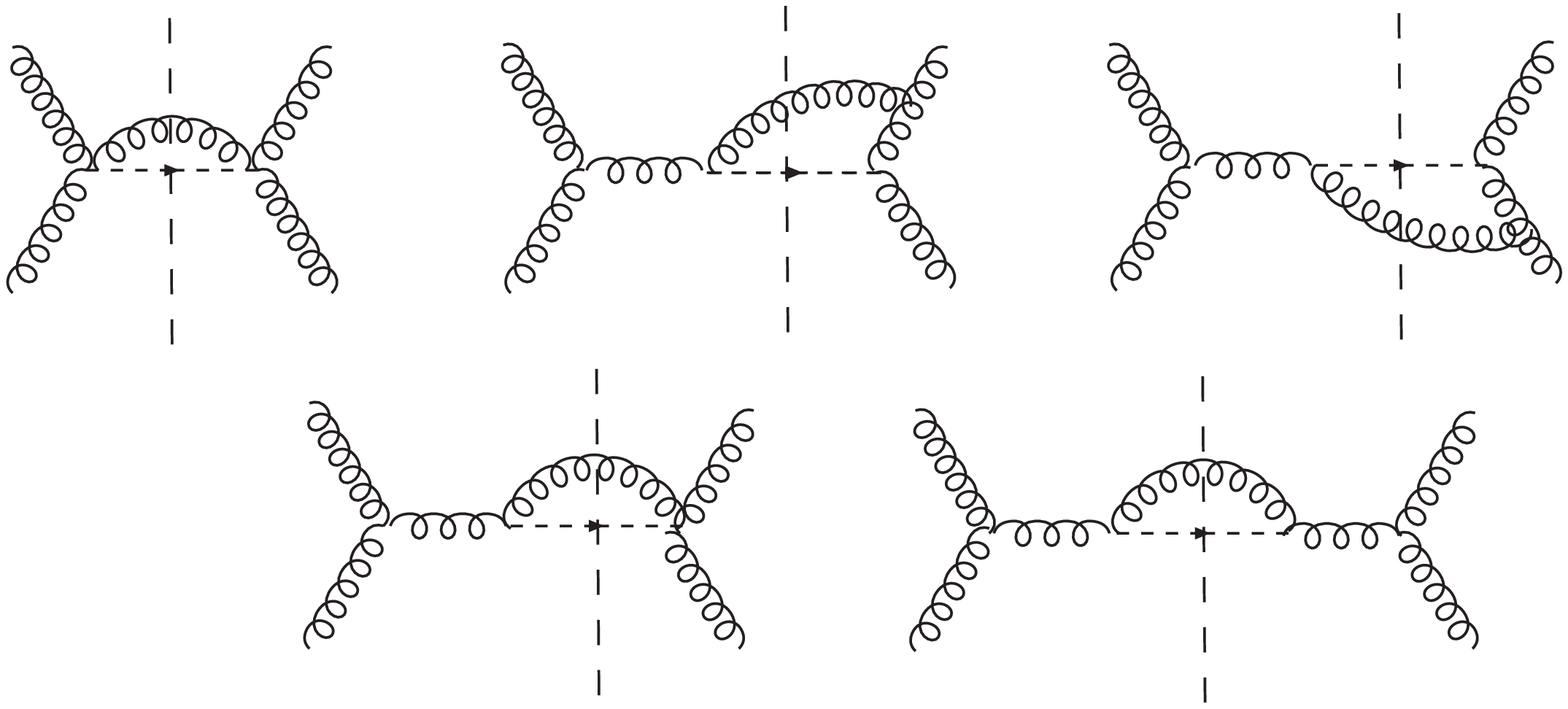}
\end{center}
\vskip -0.7cm \caption{\it Power suppressed real diagrams.}
\end{figure}

At next-to-leading order, we have both real and virtual
contributions. The real diagrams are shown in Figs.~(8-11). Figure
8 represents the contributions from the gluon radiation from
either gluons, while Fig.~9 for the interference between these
radiations. Figure 10 includes the diagrams involving the
three-gluon coupling with the scalar particle. The contribution
from Fig.~8 is
\begin{eqnarray}
\left.\frac{d\sigma}{d^2P_\perp dy}\right|_{\rm
fig.8}=\sigma_0\frac{C_A\alpha_s}{\pi^2}\frac{\mu^\epsilon}{P_\perp^2}\left\{x_1
\left[\frac{1-x_1}{x_1}+x_1(1-x_1)+\frac{x_1}{2}\right]\delta(x_2-1)
+(x_1\leftrightarrow x_2)\right\}\
\end{eqnarray}
In the above result, we have kept the pre-factor up to ${\cal
O}(\epsilon)$, because $1/P_\perp^2$ will lead to infrared
divergences when Fourier-transforming to the impact parameter
space. The contribution from Fig.~ 9 is,
\begin{eqnarray}
\left.\frac{d\sigma}{d^2P_\perp dy}\right|_{\rm
fig.9}&=&\sigma_0\frac{C_A\alpha_s}{\pi^2}\frac{\mu^\epsilon}{P_\perp^2}\left\{\left[
\left(\frac{x_1}{(1-x_1)_+}-\frac{x_1}{2}\right)+\frac{1}{2}\ln\frac{M^2}{P_\perp^2}
\delta(x_1-1)\right]\delta(x_2-1) \right. \nonumber\\
&&\left.  +(x_1\leftrightarrow x_2)\right\} \ ,
\end{eqnarray}
and the contribution from all diagrams in Fig.~10 is,
\begin{eqnarray}
\left.\frac{d\sigma}{d^2P_\perp dy}\right|_{\rm
fig.10}&=&\sigma_0\frac{C_A\alpha_s}{\pi^2}\frac{\mu^\epsilon}{P_\perp^2}
\left\{\left(-\frac{x_1^2+x_1}{2}\right)\delta(x_2-1)+(x_1\leftrightarrow
x_2) \right\}\ .
\end{eqnarray}
The diagrams in Fig. 11 are power-suppressed at low-transverse
momentum, although their contributions are leading when the
transverse momentum is on the order of $M$. The sum of the
contributions from all the real diagrams is
\begin{eqnarray}
\left.\frac{d\sigma}{d^2P_\perp dy}\right|_{\rm
real}&=&\sigma_0\frac{C_A\alpha_s}{\pi^2}\frac{\mu^\epsilon_{\rm
IR}}{P_\perp^2}\left\{\left[x_1
\left(\frac{x_1}{(1-x_1)_+}+\frac{1-x_1}{x_1}+x_1(1-x_1)\right)\right.\right.\nonumber\\
&&\left.\left.+\frac{1}{2}\ln\frac{M^2}{P_\perp^2}
\delta(x_1-1)\right]\delta(x_2-1)   +(x_1\leftrightarrow
x_2)\right\} \ .
\end{eqnarray}
The above agrees with a corresponding result in
Ref.~\cite{Kauffman:1991jt}. After Fourier-transforming to the
impact parameter space, one finds,
\begin{eqnarray}
W^{(1)}(x_i,b,M^2)|_{\rm
real}&=&\frac{\alpha_sC_A}{\pi}\left\{\left[\left(-\frac{2}{\epsilon_{\rm
IR}}-\gamma_E+ \ln\frac{4}{4\pi\mu^2b^2}
\right)\delta(x_2-1)x_1\left(\frac{x_1}{(1-x_1)_+}\right.\right.\right. \nonumber\\
&&\left.\left.+x_1(1-x_1+\frac{1-x_1}{x_1})\right)+(x_1\rightarrow
    x_2)\right]\nonumber\\
&&\left.+\delta(x_1-1)\delta(x_2-1)\left[\frac{4}{\epsilon^2_{\rm
IR}}\right.- \frac{2}{\epsilon_{\rm
IR}}\left(\ln\frac{M^2}{4\pi\mu^2}+\gamma_E\right)+\frac{1}{2}\ln^2\left(\frac{M^2}{4\pi\mu^2}\right)
    \right.\nonumber\\
    &&\left.\left.-\frac{1}{2}\ln^2
\left(\frac{M^2b^2}{4}e^{2\gamma_E}\right)
+\gamma_E\ln\frac{M^2}{4\pi\mu^2}+\frac{\gamma_E^2}{2}-\frac{\pi^2}{12}\right]\right\}\
.
\end{eqnarray}
The double and single poles above correspond to infrared
divergences only.

\begin{figure}[t]
\vskip 0.2cm
\begin{center}
\includegraphics[height=4.0cm]{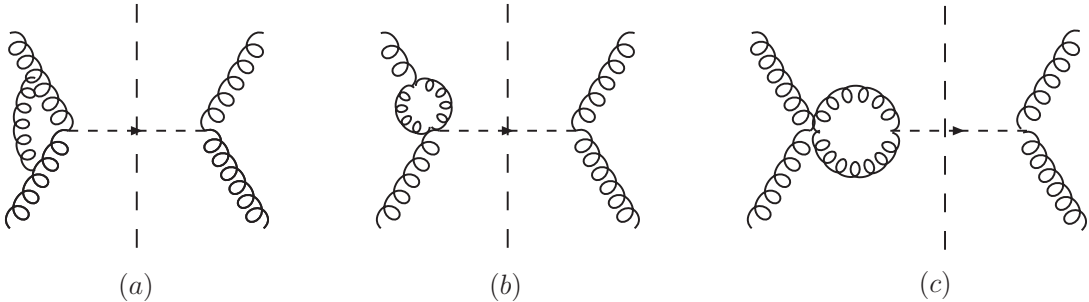}
\end{center}
\vskip -0.7cm \caption{\it Virtual diagrams contribution to the
scalar particle production through gluon fusion.}
\end{figure}

Now we turn to the virtual diagrams shown in Fig.~12. These
diagrams contribute to the cross section as
\cite{Dawson:1990zj,{Djouadi:1991tk}}
\begin{eqnarray}
W(x_i,b,M^2)|_{\rm
virtual}&=&\delta(x_1-1)\delta(x_2-1)\frac{\alpha_sC_A}{\pi}\left[-\frac{4}{\epsilon^2}
    +\frac{2}{\epsilon}\left(\ln\frac{M^2}{4\pi\mu^2}+\gamma_E\right)-
\frac{1}{2}\ln^2\left(\frac{M^2}{4\pi\mu^2}\right)\right.\nonumber\\
    &&\left.-\gamma_E\ln\frac{M^2}{4\pi\mu^2}
    -\frac{\gamma_E^2}{2}+\frac{7\pi^2}{12}\right]\
    \label{hvirtual}
\end{eqnarray}
The $1/\epsilon^2$ pole above cancels that from the real diagrams.
The above result has ultra-violet divergences which comes from the
composite gluon operator in the effective coupling. This
divergence can be cancelled by a charge renormalization
contribution,
\begin{eqnarray}
W(x_i,b,M^2)|_{\rm
charge}=\frac{\alpha_sC_A}{\pi}2\beta_0\left(\frac{M^2}{4\pi\mu^2}\right)^{-\epsilon/2}
\left(-\frac{2}{\epsilon}+\gamma_E\right)\delta(x_1-1)\delta(x_2-1)\
.
\end{eqnarray}
where the extra factor is introduced so that renormalization scale
is set at $\mu_{UV}^2  = M^2$. The gluon wave function
renormalization constant $Z_A=1$ to all orders in perturbation
theory in the present scheme. By summing up all the contributions,
we get $W(x_i,b,M^2)$ at one-loop order,
\begin{eqnarray}
W^{(1)}(x_i,b,Q^2)&=&\frac{\alpha_sC_A}{\pi}\left\{\left[x_1{\cal
P}_{gg}(x_1)\delta(x_2-1)\left(-\frac{2}{\epsilon}-\gamma_E+\ln\frac{4}{4\pi\mu^2b^2}
    \right)+(x_1\rightarrow x_2)\right]\right.\nonumber\\
    &&+\left.\delta(x_1-1)\delta(x_2-1)
\left[2\beta_0\ln\frac{b^2M^2}{4}e^{2\gamma_E}-\frac{1}{2}\ln^2\left(\frac{M^2b^2}{4}e^{2\gamma_E}\right)
    +\frac{\pi^2}{2}\right]\right\}\ .\nonumber\\\label{higgstotal}
\end{eqnarray}
The soft divergences have been cancelled between real and virtual
diagrams, and there are remaining collinear divergences.

Now we can verify the factorization formula in Eqs.~(\ref{fac1}),
(\ref{fac2}), using the one-loop results in Eqs.~(\ref{tmdgd}),
(\ref{softh}). After the subtractions, we obtain the hard factor
at one loop,
\begin{eqnarray}
H^{(1)}(M^2,\mu^2,\rho)&=&\frac{\alpha_sC_A}{\pi}\left[\ln\frac{M^2}{\mu^2}
    \left(2\beta_0+\frac{1}{2}\ln\rho^2-\frac{3}{2}\right)-
    \frac{3}{4}\ln\rho^2+\frac{1}{8}\ln^2\rho^2+\pi^2+\frac{7}{2}\right]\
    ,\nonumber\\\label{hardh}
\end{eqnarray}
where a special coordinate system has been chosen:
$x_1^2\zeta_1^2=x_2^2\zeta_2^2=\rho M^2$. Not only the collinear
divergences in the cross section have been cancelled by the gluon
distributions, the $b$-dependence has also been entirely isolated
in the parton distributions and soft factor. The hard factor $H$
depends only on the hard scale $M^2$, the factorization scale
$\mu$, and the parameter $\rho$. Thus, the factorization
conjectured in Eq.~(\ref{fac1}) is verified to one-loop order.
Furthermore, according to the general arguments provided in
\cite{ColSop81,JiMaYu04}, all-order QCD corrections to
scalar-particle production can be factorized in a similar way.

\subsection{Factorization at Small $b\ll 1/ \Lambda_{\rm QCD}$}

When $b$ is small, there is the standard QCD factorization in
terms of Feynman parton distributions for $W(x_i,b,M^2)$
\cite{ColSopSte85}. In this subsection, we will try to recover
this factorization, using the results in the last subsection.

According to \cite{ColSopSte85}, the cross section $W(x_i,b,M^2)$
at small $b$ can be written as,
\begin{eqnarray}
W(x_i,b,M^2)=\int\frac{d\xi_1}{\xi_1}\frac{d\xi_2}{\xi_2}\xi_1g(\xi_1,\mu)
\xi_2g(\xi_2,\mu) C_{gg}(\frac{x_1}{\xi_1},b,\mu,M^2)
C_{gg}(\frac{x_2}{\xi_2},b,\mu,M^2) \ ,
\end{eqnarray}
where $g(x,\mu)$ is the usual gluon distribution, and the
coefficient functions $C_{gg}$ is perturbative. They both depend
on the factorization scale $\mu$, but the dependence cancels out
in the product.

Using $W(x_i,b,M^2)$ from the last subsection, we can calculate
the $C_{gg}$ to one-loop. The perturbative gluon distribution up
to one-loop is
\begin{eqnarray}
g(x)=\delta(x-1)+\frac{\alpha_sC_A}{\pi}{\cal
P}_{gg}(x)\left(-\frac{2}{\epsilon}+\gamma_E+\ln\frac{1}{4\pi}\right)\
.
\end{eqnarray}
Subtracting the above from the one-loop $W(x_i,b,M^2)$, we have
\begin{eqnarray}
C_{gg}(x,b,\mu,M^2)&=&\delta(x-1)+\frac{\alpha_sC_A}{\pi}\left\{x{\cal
P}_{gg}(x)\ln\left(\frac{4}{b^2\mu^2}e^{-2\gamma_E}\right)\right.\nonumber\\
&&\left.+\delta(x-1)\left[\beta_0\ln\left(\frac{b^2M^2}{4}e^{2\gamma_E}\right)-
\frac{1}{4}\ln^2\left(\frac{M^2b^2}{4}e^{2\gamma_E}\right)+\frac{\pi^2}{4}\right]\right\}
\ .
\end{eqnarray}
The $\mu$-dependence of the $C_{gg}$ function is controlled by the
same evolution equation as for the gluon distribution, with the
opposite sign. When $b^2M^2$ is large, one needs to re-sum the
large double logarithms to all orders in perturbation theory. This
is difficult to do systematically in the present formalism.
However, this can be done straightforwardly in terms of the TMD
factorization in the previous subsection.

\section{Resummation of Large Double Logarithms}

The one-loop result for scalar-particle production shows that
there exists large double logarithms in the hard scale $M^2$
(e.g., see Eq.~(\ref{higgstotal})). These large logarithms appear
at every order of the perturbative expansion, $\alpha_s^n\ln^{2m}
M^2/P_\perp^2$ where $m\leq n$ and $P_\perp$ is the transverse
momentum. In order to make reliable predictions, one has to re-sum
these large logarithms. In this section, we will follow the
Collins-Soper-Sterman (CSS) framework \cite{ColSopSte85} to
perform the resummation. The results can be used as a double check
of the resummation studies in the literature for the standard
model Higgs production at hadron colliders
\cite{Hinchliffe:1988ap,{Kauffman:1991jt},{Yuan:1991we}}.

\subsection{Collins-Soper-Sterman Resummation at Small b}

According to the QCD factorization, the Fourier transformation of
the cross section for scalar-particle production is
\begin{eqnarray}
W(x_i,b,M^2)=x_1g(x_1,b,\mu,\rho M^2,\rho) x_2g(x_2,b,\mu,\rho
M^2,\rho)S(b,\mu,\rho)H(M^2,\rho,\mu) \
\end{eqnarray}
In the above equation we have set
$x_1^2\zeta_1^2=x_2^2\zeta_2^2=\rho M^2$ in a special frame of
coordinates. The $M^2$ dependence of $W$ can be studied through
the following differential equation \cite{ColSopSte85},
\begin{equation}
\frac{\partial W(x_i,b,M^2)}{\partial \ln M^2}=(K+G') W(x_i,b,M^2)
\ ,
\end{equation}
where $K$ and $G'$ are soft and hard evolution kernels. The soft
part $K$ depends on the scale $1/b^2$ and the renormalization
scale $\mu$, while $G'$ depends on the hard scale $M^2$ and $\mu$.
$K$ can be calculated from the Collins-Soper evolution equation
for the TMD gluon distributions discussed in Sec.IID. $G'$
contains the contributions from the gluon distribution as well as
from the hard factor $H$. From the result of the previous
sections, the sum of $K$ and $G'$ at one-loop order is,
\begin{equation}
K+G'=-\frac{\alpha_sC_A}{\pi}\ln
\left(\frac{M^2b^2}{4}e^{2\gamma_E-2\beta_0}\right) \ ,
\end{equation}
where the $\rho$ dependence between various terms cancels out. Our
previous result for $K$ at one loop is
\begin{equation}
K=K_g=-\frac{\alpha_sC_A}{\pi}\ln\frac{b^2\mu^2}{4}e^{2\gamma_E} \
, \label{kh}
\end{equation}
which is valid when $b$ is small enough. At large $b$, the
perturbative expansion breaks down. The hard part $G'$ is,
\begin{equation}
G'=-\frac{\alpha_sC_A}{\pi}\left(\ln
\frac{M^2}{\mu^2}-2\beta_0\right) \ . \label{gh}
\end{equation}
Both the soft and hard parts $K$ and $G'$ also obey the
renormalization group equation \cite{ColSopSte85},
\begin{equation}
\frac{\partial K}{\partial \ln\mu}=-\frac{\partial G'}{\partial
\ln \mu}=-\gamma_{Kg} \ .
\end{equation}
Resummation is done here by solving the above equations.

Integrating over $\ln M^2$ and $\ln \mu^2$, one finds
\cite{ColSopSte85}
\begin{eqnarray}
W(x_i,b,M^2)=e^{-{\cal S}_{Sud}(M^2,b,C_1/C_2)}
W(x_i,b,C_1^2/C_2^2/b^2) \ ,
\end{eqnarray}
where the Sudakov form factor is
\begin{equation}
{\cal S}_{Sud}=\int_{C_1^2/b^2}^{C_2^2M^2}\frac{d
\mu^2}{\mu^2}\left[\ln\left(\frac{C_2^2M^2}{\mu^2}\right)
A(C_1,\mu)+B(C_1,C_2,\mu) \right]\ .
\end{equation}
Here $C_1$ and $C_2$ are two parameters of order one. The
functions $A$ and $B$ can be expanded perturbatively $\alpha_s$,
$A=\sum\limits_{i=1}^\infty
A^{(i)}\left(\frac{\alpha_s}{\pi}\right)^i$ and
$B=\sum\limits_{i=1}^\infty
B^{(i)}\left(\frac{\alpha_s}{\pi}\right)^i$. They are related to
the Collins-Soper evolution parameters by the following equation,
\begin{eqnarray}
A(C_1,\mu)&=&\frac{1}{2}\gamma_{Kg}(\mu)+\frac{1}{2}\beta\frac{\partial}{\partial
g}K(C_1,g(\mu))\ , \nonumber\\
B(C_1,C_2,\mu)&=&-K(C_1,g(\mu))-G'(1/C_2,g(\mu)) \ .
\end{eqnarray}
From Eqs.~(\ref{kh},\ref{gh}), we get the first expansion of $A$
and $B$ functions,
\begin{eqnarray}
A^{(1)}=C_A, ~~~
B^{(1)}=C_A\left[\ln\left(\frac{C_1^2}{4C_2^2}e^{2\gamma_E}\right)-2\beta_0\right]
\ .
\end{eqnarray}
To  complete the resummation, we use the factorization of
$W(x_i,b,C_1^2/C_2^2/b^2)$ in terms of the integrated parton
distributions discussed in Sec.~IVB,
\begin{eqnarray}
W\left(x_i,b,\frac{C_1^2}{C_2^2b^2}\right)&=&\int\frac{d\xi_1}{\xi_1}\frac{d\xi_2}{\xi_2}
C_{gg}\left(\frac{x_1}{\xi_1},b;\frac{C_1^2}{C_2^2b^2},\mu^2\right)
C_{gg}\left(\frac{x_2}{\xi_2},b;\frac{C_1^2}{C_2^2b^2},\mu^2\right)\nonumber\\
&& \times \xi_1 g(\xi_1,\mu) \xi_2 g(\xi_2,\mu) \ , \label{cgluon}
\end{eqnarray}
where $C$ functions also have perturbation expansion in terms of
$\alpha_s$, $C=\sum\limits_{i=0}^\infty
C^{(i)}\left(\frac{\alpha_s}{\pi}\right)^i$. From the results in
Sec.~IVB, we get the first two terms of $C_{gg}$,
\begin{eqnarray}
C_{gg}^{(0)}&=&\delta(x-1)\nonumber\\
C_{gg}^{(1)}&=&C_A\left\{x{\cal
P}_{gg}(x)\ln\left(\frac{4}{C_1^2}e^{-2\gamma_E}\right)\right.\nonumber\\
&&~~~\left.+\delta(x-1)\left[\beta_0
\ln\left(\frac{C_1^2}{4C_2^2}e^{2\gamma_E}\right)-\frac{1}{4}\ln^2\left(\frac{C_1^2}{4C_2^2}e^{2\gamma_E}\right)
+\frac{\pi^2}{4}\right]\right\}\ .
\end{eqnarray}
It is easy to check that the above $C$ function for the
gluon-gluon term can also be calculated from the following
equation \cite{JiMaYu04}
\begin{equation}
C_{gg}(x,b;\frac{C_1^2}{C_2^2b^2},\mu^2)=x\tilde C_{gg}\left(x,
b^2, \mu_g^2, \mu^2,\rho \frac{C_1^2}{C_2^2b^2}, \rho\right)\times
\sqrt{S(b,\mu_g^2,\rho)H(\frac{C_1^2}{C_2^2b^2},\mu_g^2,\rho)} \ ,
\end{equation}
with $\tilde C_{gg}$ from Eq.(\ref{cggt}), and $S$ and $H$ from
Eqs.~(\ref{softh}), (\ref{hardh}) respectively. Similarly, the
contribution from the quark distribution in the factorization form
of Eq.~(\ref{cgluon}) can also be calculated using the above
equation with $\tilde C_{g/q}$ in Eq.~(\ref{cqg}). Since $\tilde
C_{g/q}^{(0)}=0$, it is straightforward to get the expansion of
$C_{g/q}$ up to one-loop order,
\begin{eqnarray}
C_{g/q}^{(0)}&=&0\nonumber\\
C_{g/q}^{(1)}&=&\frac{C_F}{2}x\left[\frac{1+(1-x)^2}{x}\ln
\left(\frac{4}{b^2\mu^2}e^{-2\gamma_E}\right) +x\right] \ .
\end{eqnarray}
Our final resummation result is the combination of Eqs. (66),
(67), (70), (71) and (73).

\subsection{Comparison with Previous Calculations}

In the literature, the above coefficient functions have also been
calculated by various authors
\cite{{Catani:1988vd},Hinchliffe:1988ap,{Kauffman:1991jt},{Yuan:1991we}}.
They are normally extracted from the comparison between the
fixed-order results and the expansion of the resummation formula,
and expressed with the so-called canonical parameters, with
$C_1=2e^{-\gamma_E}$ and $C_2=1$. Choosing these specific values,
our results for $A$, $B$, $C$ at one-loop order are,
\begin{equation}
A^{(1)}=C_A,~~~B^{(1)}=-2\beta_0C_A,~~~C_{gg}^{(1)}=C_A\frac{\pi^2}{4}\delta(x-1)
\ .
\end{equation}
$A^{(1)}$ and $B^{(1)}$ agree with the previous calculations
\cite{{Catani:1988vd},Hinchliffe:1988ap,{Kauffman:1991jt}}, while
$C^{(1)}$ lacks an additional term $11/4$. The difference comes
from the effective coupling between the Higgs boson and gluons
$g_\phi$ \cite{Dawson:1990zj,{Djouadi:1991tk}}
\begin{equation}
g_{H}=-\frac{1}{3{\cal
V}}\frac{\alpha_s}{\pi}\left[1+\frac{11}{4}\frac{\alpha_s}{\pi}+\cdots\right]\
,
\end{equation}
where ${\cal V}$ is the vacuum expectation value of the Higgs
field. The second term will contribute at one-loop order in
$\alpha_s$ and modify the result of Eq.~(\ref{hvirtual}). Taking
this into account, our results coincide with the previous
calculations \cite{Hinchliffe:1988ap,{Kauffman:1991jt}}.

In \cite{Yuan:1991we}, the full of dependence on $C_1$ and $C_2$
of the coefficient functions were also calculated for the standard
model Higgs boson production. Our result on $B^{(1)}$ agrees with
theirs, while $C^{(1)}$ does not \footnote{After correcting some
errors in \cite{Yuan:1991we}, these two calculations agree on
$C^{(1)}$.}.

\section{Conclusion}

In this paper, we have studied scalar-particle production at
one-loop order to examine the factorization of the gluon initiated
semi-inclusive processes at hadron colliders. We introduced and
calculated the gauge-invariant TMD gluon distribution at one-loop
order, and established their connections with the integrated
parton distributions at small $b$. We verified the factorization
theorem at one-loop order, where the cross section can be
decomposed into products of TMD parton distributions, soft and
hard factors. The large logarithms in the cross section were
resummed following the CSS formalism, and the coefficient
functions were obtained. The factorization to all orders in QCD
coupling is assumed following the general arguments in
\cite{ColSop81,JiMaYu04}.

The present study can be easily extended to the polarized gluon
TMD distributions and their contributions to the spin asymmetries
for the semi-inclusive processes. Scalar-particle production can
also be generalized to other production processes, e.g.,
heavy-quark pairs and heavy quarkonium production, di-photon and
di-hadron and photon-hadron production, and di-jet correlations.
We will carry out further studies in future publications.

The present study of factorization for semi-inclusive processes
can be extended to two-loop order, which can be used to compare
with the NNLO calculation of the standard model Higgs production.
It will be interesting to calculate the resummation coefficient
functions from the present approach, and compare with the previous
results obtained by the expansion method \cite{deFlorian:2000pr}.

As a final remark, we consider the small-$x$ property of the gluon
TMD distributions. From the results in Sec.II, we found that the
TMD gluon distribution at small $x$ has the similar behavior as
the integrated gluon distribution, where the small $x$ resummation
is usually necessary. The so-called BFKL evolution equation
\cite{Lipatov:1976zz} is usually used to make the resummation. It
will be interesting to examine whether the TMD gluon distributions
obey such evolution equation at small $x$ \cite{Sterman:1995fz},
and to explore the connection between the present approach and the
so-called $k_t$-factorization approach \cite{Col91}.
Phenomenologically, the small-$x$ effects are important for the
Higgs boson and vector-boson production at LHC
\cite{Berge:2004nt}.

\section*{Acknowledgments}
F.Y. thanks W. Vogelsang for the collaboration at the early stage
of this work. We thank C.P. Yuan for the communication about the
results of \cite{Yuan:1991we}. X. J. was supported by the U. S.
Department of Energy via grant DE-FG02-93ER-40762. J.P.M.  and
also X.J. are supported by National Natural Science Foundation of
P.R. China. F.Y. is grateful to RIKEN, Brookhaven National
Laboratory and the U.S. Department of Energy (contract number
DE-AC02-98CH10886) for providing the facilities essential for the
completion of his work.

\end{document}